\documentclass{aastex62}
    
\usepackage[utf8]{inputenc}
\usepackage{graphicx}
\usepackage{bm}
\usepackage{color}      
\usepackage{amsmath}
\usepackage{natbib}

\begin{document}        
   
\title{Simulation of plasma emission in magnetized plasmas}

\author{Sang-Yun Lee}
\email{sangyun.lee@mail.wvu.edu}
\affiliation{Department of Physics and Astronomy, West Virginia University, Morgantown, WV 26506}

\author{Peter H. Yoon}
\email{yoonp@umd.edu}
\affiliation{Institute for Physical Science and Technology, University of Maryland, College Park, MD 20742-2431}

\author{Ensang Lee}
\affiliation{School of Space Research, Kyung Hee University, Yongin, Gyeonggi 17104, Republic of Korea}

\author{Weichao Tu}
\affiliation{Department of Physics and Astronomy, West Virginia University, Morgantown, WV 26506}

\date{\today}

\begin{abstract}
The recent Parker Solar Probe (PSP) observations of type III radio bursts show that the effects of finite background magnetic field can be an important factor in the interpretation of data. In the present paper, the effects of background magnetic field on the plasma emission process, which is believed to be the main emission mechanism for solar coronal and interplanetary type III radio bursts, are investigated by means of the particle-in-cell simulation method. The effects of ambient magnetic field is systematically surveyed by varying the ratio of plasma frequency to electron gyro-frequency. The present study shows that for a sufficiently strong ambient magnetic field, the wave-particle interaction processes lead to a highly field-aligned longitudinal mode excitation and anisotropic electron velocity distribution function, accompanied by a significantly enhanced plasma emission at the second harmonic plasma frequency. For such a case, the polarization of the harmonic emission is almost entirely in the sense extraordinary mode. On the other hand, for moderate strengths of the ambient magnetic field, the interpretation of the simulation result is less than clear. The underlying nonlinear mode coupling processes indicate that to properly understand and interpret the simulation results require sophisticated analyses involving interactions among magnetized plasma normal modes including the two transverse modes of the magneto-active plasma, namely, extraordinary and ordinary modes, as well as electron-cyclotron-whistler, plasma oscillation, and upper-hybrid modes. At present, a nonlinear theory suitable for quantitatively analyzing such complex mode-coupling processes in magnetized plasmas is incomplete, which calls for further theoretical research, but the present simulation results could provide a guide for future theoretical efforts.
\end{abstract}
    
 

\section{Introduction}

Solar radio burst phenomena in the meter wavelengths were discovered in the 1950s and subsequently classified into five categories \citep{WildMcCready50/09, Wild50/09, Wild50/12, Wild+54/09}, of which, the present paper is concerned with some fundamental theoretical aspects undergirding the type III radio bursts. For an excellent review on the subject matter of solar radiophysics until the decade of mid 1980s, see the collection of articles in the monograph edited by \citet{McLeanLabrum1985}. For a more recent review on the subject of solar type III bursts, see, e.g. \citet[and references therein]{ReidRatcliffe2014}. The general theoretical framework most commonly adopted in the literature for type III radio bursts is based upon the pioneering work of \citet{GinzburgZheleznyakov58/10}, which in the more modern form, can be described as the plasma emission mechanism based on the weak turbulence paradigm. The plasma weak turbulence theory \citep{Kadomtsev, Sagdeev, Tsytovich1970, Davidson, Sitenko, Melrose1980} found in the literature is of two varieties. One is based upon the intuitive method of semi-classical (or semi-quantum mechanical) approach, while the other takes the classical non-equilibrium statistical mechanical approach. For a systematic discourse on the plasma emission theory within the framework of semi-classical formulation of weak turbulence theory, see the excellent monograph by \citet{Melrose1980}. The semi-classical theory was pioneered by scientists from the former Soviet Union, most notably, by \citet{Tsytovich1970}. The statistical method is also available in the literature, e.g., see \citep{Kadomtsev, Davidson, Sitenko, Yoon00/12, Yoon06/02, Yoon+12/10, Yoon2019}. The non-equilibrium statistical mechanics, or equivalently, kinetic theory starts from a governing equation of the many-particle system, and systematically introduce perturbation expansion and statistical averages. The semi-classical and statistical theories in the end, produce equivalent set of equations, which can be applied for the detailed investigations of solar type III (and type II) radio bursts.

According to the weak turbulence-based standard theory of plasma emission, active flare events in the solar chromosphere generate bursts of energetic electrons that travel outward along open magnetic field lines and interact with the background plasma via a beam-plasma instability. This bump-on-tail instability excites the longitudinal electrostatic Langmuir waves ($L$), which subsequently undergo nonlinear three-wave decay/coalescence and nonlinear wave-particle scattering, to generate the backward traveling $L$ mode mediated by the generation of low-frequency ion-acoustic ($S$) mode. The merging of the forward- and backward-propagating $L$ modes lead to the radiation emission at twice the plasma frequency (harmonic, or $H$, emission). The beam-generated $L$ mode also decays into an $S$ mode and a transverse electromagnetic mode ($T$) at the plasma frequency. This process, which is equivalent to the transformation of longitudinal mode to radiation, is responsible for the generation of radiation at the fundamental ($F$) plasma frequency. This standard scenario has been investigated for many decades and various aspects of the plasma emission process have been studied in the literature \citep{Tsytovich67, KaplanTsytovich68/06, Melrose82/04, GoldmanDuBois82/06, Goldman83/12, Melrose87/03, Cairns87/10a, RobinsonCairns98/08a, RobinsonCairns98/08b, Zlotnik+98/03, Kontar01/04, Kontar01/08b, Kontar01/08c, Gosling+03/07, Li+05/01, Li+05/05, Li+06/09, Li+08/06, Li+11/03, Li+12/07, Ratcliffe+12/12, ReidKontar17/02, Krafft, Ziebell+14/11, Ziebell+15/06, Ziebell+16/02, Tkachenko}.

The weak turbulence-based plasma emission is the most widely-accepted theoretical framework for interpretation of solar type III (and type II) radio bursts. However, there are alternative theories as well. For instance, the dipole radiation emanating from the solitary electrostatic structure, which is a consequence of the so-called strong Langmuir turbulence theory \citep{Zakharov1972}, is an alternative radiation process \citep{Papadopoulos1974, Papadopoulos1978, Smith1979, Goldstein1979, Goldman1980, Goldman1984, Robinson1997, Thejappa2013, Thejappa2020, Che2017}. The radiation mechanism based on the strong turbulence processes may be particularly applicable for intense radio bursts. For the fundamental emission, linear mode conversion process \citep{Hinkel-Lipsker1992, Kim2007} may also be a significant aspect that cannot be overlooked when interpreting spacecraft data, which includes the recent Parker Solar Probe data \citep{Krasnoselskikh2019, VolokitinKrafft}.

The standard weak turbulence theory has successfully addressed the plasma emission problem, as noted above. Recently, complete equations of electromagnetic weak turbulence theory are solved in order to quantitatively investigate the plasma emission process \citep{Ziebell+15/06}, and its validity was confirmed by 2+1/2 dimensional (two dimensional space and three dimensional velocity) particle-in-cell (PIC) simulation \citep{Lee19} -- see also, \citep{Ratcliffe+14/12}. Note that PIC simulations of plasma emission process have been carried out, as evidence by a number of publications in the literature \citep{Kasaba+01/09, Umeda, KarlickyVandas07/12, Rhee+09/01, Rhee+09/03, Ganse+12/06, Ganse+12/10, ThurgoodTsiklauri15/12, Che2017, Henri, Li, Krafft2021}. It should be noted, however, that a direct and quantitative comparison between the PIC simulation and weak turbulence theory is rarely done \citep{Ratcliffe+14/12, Lee19}.

A common thread that connects various works regarding the standard model of the plasma emission theory, and also the existing PIC simulations thereof, is that the effects of background quasi static magnetic field is ignored. The assumption of unmagnetized plasma is valid in the interplanetary space sufficiently far away from the solar surface, since the solar wind magnetic field decreases quite rapidly as one moves away from the Sun. However, near the solar active region, which is associated with the type III radio source, the effects of strong magnetic field can be significant \citep{Morosan}. The contemporary Parker Solar Probe (PSP) measurement of type III radio bursts also leads to open questions regarding the polarization properties \citep{Pulupa}. Without the ambient magnetic field the type III bursts should be unpolarized, but \citet{Pulupa} report circular polarizations, which indicate that the type III emissions occurring close to the Sun is influenced by the background magnetic field. In a similar vein, \citet{Ma} analyzed the PSP data by surveying the low frequency cutoffs associated with the type III bursts and speculate that the cyclotron maser emission may also be effective in the radiation process. Further, \citet{Chen+2021} modeled the interplanetary type IIIb emissions observed by PSP by means of the cyclotron maser instability mechanism. The cyclotron maser emission mechanism, which requires the presence of a strong ambient magnetic field and a loss-cone feature (or perpendicular population inversion in momentum space) associated with the energetic electrons, was first suggested as an alternative mechanism for type III radio bursts near the source region by \citet{Wu}, but the recent PSP measurements revived the interest in such an alternative mechanism \citep{Chen+2021, Zhou+2020}.

In another development, \citet{Ni20, Ni21} analyzed the possibility of plasma emission by mode coupling among the cold plasma modes in magnetized plasmas, in particular, the emission involving whistler (or electron cyclotron) mode waves. In the literature, the plasma emissions in magnetized plasmas have been investigated in some detail by \citet{Trakhtengerts, MelroseSy, Melrose78, Melrose80, Zlotnik, Willes}, but in these works, the standard plasma emission mechanism developed for unmagnetized plasmas is extended to the magnetized case, where high-frequency waves undergo decay interaction with low-frequency sonic type of modes (such as the magneto-acoustic mode, which is a thermal mode that does not propagate in cold plasmas), but \citet{Ni20, Ni21} revived the old idea of plasma emission in magnetized plasmas involving the whistler mode type of waves, which does not require a finite-temperature effects. Such an idea was proposed earlier by \cite{Chiu, Chin}, which was subsequently criticized by \citet{Melrose75}, but \citet{Ni20, Ni21} re-investigated such a possibility by means of modern PIC simulation.

The contemporary developments and research activities as outlined above show that the venerable plasma emission theory needs to be extended to include the effects of ambient magnetic field. While some early works \citep{Trakhtengerts, MelroseSy, Melrose78, Melrose80, Zlotnik, Willes} have indeed attempted to address such an issue, the problem is far from complete. Unlike the unmagnetized situation, where the standard weak turbulence theory has advanced to a mature stage so that a direct numerical comparison with PIC simulation is possible \citep{Lee19}, the situation for the magnetized case is at an infantile stage in comparison. The complete self-consistent set of equations that readily lend themselves to numerical (or theoretical) analysis are not available at this time.

The purpose of the present paper is to address the influence of ambient magnetic field on plasma emission by carrying out the particle-in-cell (PIC) simulation. The present paper performs PIC simulations of electron beam-plasma instability in a magnetized plasma and the ensuing nonlinear mode coupling processes, which includes the radiation emissions at the fundamental plasma frequency and/or its harmonic. The PIC simulations are performed by varying the degree of magnetization in the plasma. This is accomplished by choosing differing ratios of plasma-to-gyro frequencies, $\omega_{pe}/\Omega_{ce}$, where $\omega_{pe}=(4\pi n_0e^2/m_e)^{1/2}$ is the plasma oscillation frequency, and $\Omega_e=eB_0/m_ec$ is the electron cyclotron (or gyro) frequency. When the ratio $\omega_{pe}/\Omega_{ce}$ is high, then the plasma is weakly magnetized, and {\it vice versa}. By carrying out the simulation with different values of $\omega_{pe}/\Omega_{ce}$ we aim to establish a constraint against which, future quantitative theories of plasma emission in magnetized plasma can be tested.

With this aim, we present our findings, which is organized as follows: In section \ref{sec2} a brief description of the simulation code is presented. Then, in section \ref{sec3} results from the PIC simulations, together with the interpretations, are presented. Finally, in section \ref{sec4} we summarize the major findings and present our conclusions.

\section{Particle-in-Cell Simulation of Plasma Emission in Magnetized Plasma}\label{sec2}

The EM particle-in-cell simulation adopted in the present paper is briefly described herewith. The numerical program is a two-dimensional relativistic and electromagnetic particle-in-cell (PIC) simulation computer code. The simulation box size is $L_X=64.76 c/\omega_{pe}$ for all 4 cases and $L_Y=L_X=64.76 c/\omega_{pe}$ for $\omega_{pe}/\Omega_{ce}=100$ and 10, but $L_Y=2L_X=129.53$ for $\omega_{pe}$/$\Omega_{ce}=3$ and 1 for higher resolution of $k_{\perp}$, where $c$ is the speed of light, $V_{th}$ is the electron thermal speed and $\omega_{pe}$ is the electron plasma frequency. The number of grids is $N_X \times N_Y=1024\times1024$ for $\omega_{pe}/\Omega_{ce}=100$ and 1, and $N_X \times N_Y = 1024 \times 2048$ for $\omega_{pe}/\Omega_{ce}=3$ and 1, so that the grid size is $\Delta x = \sqrt{2} \lambda_D$. The simulation time step is $\Delta t=\Delta x /2c$, which satisfies the Courant-Friedrichs-Lewy (CFL) condition, where $\Delta t$ is the time step. 

The number of particles is 200 per grid per species, and we used three kinds of species: protons, background electrons, and beam electrons. The proton-to-electron mass ratio is realistic, $m_p/m_e=1836$, where $m_p$ is the proton mass and $m_e$ is the electron mass. The electron thermal speed is $V_{th}^2/c^2=4.0\times10^{-3}$. The electron temperature is 7 times higher than proton temperature, $T_i/T_e=1/7$, where $T_i$ is the ion (proton) temperature and $T_e$ is the electron temperature. The ambient magnetic field is directed along $X$ axis.

The beam consists of $0.1\%$ of the total electron content, namely, $n_b/n_0=10^{-3}$, where $n_b$ is the beam number density and $n_0$ is the electron number density. The Maxwellian beam temperature is the same as the background electron temperature, $T_b=T_e$, where $T_b$ is the electron beam temperature. The plasma-to-cyclotron frequency ratio is systematically varied from $\omega_{pe}/\Omega_{ce}=100$, to 10, to 3, to 1, where $\Omega_{ce}$ is electron cyclotron frequency. The boundary conditions are periodic for both $X$- and $Y$-directions.

The above described parameters are chosen as in the paper by \citet{Ziebell+15/06}. In all our simulations where we adopted the same average beam drift speed, $V_b/v_{th}=6$. Again, such choices reflects the same initial condition as those considered by \citet{Ziebell+15/06}. The PIC code simulation is initiated with the distributions of electrons and protons that preserve the current-free condition so that background electrons have a slight negative drift against the background of protons. The initial velocity distributions for the electrons is specified according to
\begin{eqnarray}
F_e({\bf V},0) &=& \left(1-\frac{n_b}{n_0}\right)
\left(\frac{m_i}{2\pi T_e}\right)^{3/2}
\exp\left(-\frac{m_eV_\perp^2}{2T_e}
-\frac{m_e(V_\parallel+V_0)^2}{2T_e}\right)
\nonumber\\
&& +\frac{n_b}{n_0}\left(\frac{m_e}{2\pi T_b}\right)^{3/2}
\exp\left(-\frac{m_eV_\perp^2}{2T_b}
-\frac{m_e(V_\parallel-V_b)^2}{2T_b}\right).
\end{eqnarray} 

\section{Numerical Results}\label{sec3}

\begin{figure}
\centering
\includegraphics[width=0.75\textwidth]{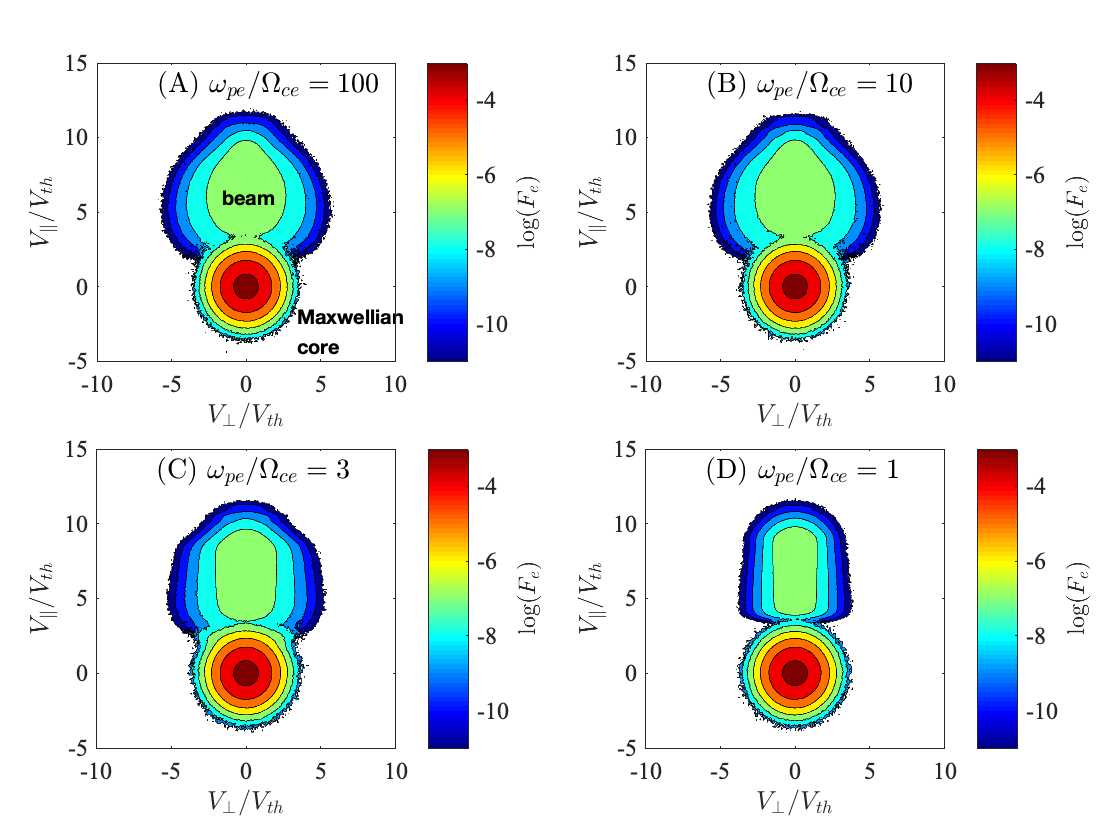}
\caption{Electron velocity distribution function at $\omega_{pe}t=2000$ for (A) $\omega_{pe}/\Omega_{ce}=100$; (b) $\omega_{pe}/\Omega_{ce}=10$; (C) $\omega_{pe}/\Omega_{ce}=3$; and (D) $\omega_{pe}/\Omega_{ce}=1$.}
\label{F1}
\end{figure}

We begin by presenting the two-dimensional phase space plot of the electron velocity distribution function (VDF) at $\omega_{pe}t=2000$. Figure \ref{F1} shows four difference cases, (A) $\omega_{pe}/\Omega_{ce}=100$; (B) $\omega_{pe}/\Omega_{ce}=10$; (C) $\omega_{pe}/\Omega_{ce}=3$; and (A) $\omega_{pe}/\Omega_{ce}=1$. Case (A) is almost identical to the unmagnetized situation \citep{Lee19} in that the Maxwellian core is not much affected by the beam-plasma instability, while the beam electrons are shown to have undergone not only the plateau formation along $V_\parallel$, but also a perpendicular spreading along $V_\perp$. Note that the present PIC simulation, as well as that of \citet{Lee19}, is carried out over fully three-dimensional velocity space, but the spatial dimension is in 2D space. In comparison, the quasilinear calculation carried out by \citet{Ziebell+15/06} was done in 2D velocity space as well as in 2D wave number space. Recent papers by \citet{Harding, Melrose2021} discuss some issues associated with the velocity space diffusion in 1D versus 3D configuration. The quasilinear calculation by \citet{Ziebell+15/06} (not shown in the present paper) is the intermediate case of 2D, which upon comparison with the simulated VDF in 3D in the paper by \citet{Lee19}, was shown to be in good qualitative agreement.

Panel (B), which is for $\omega_{pe}/\Omega_{ce}=10$ shows only a minimal difference when compared with the case for $\omega_{pe}/\Omega_{ce}=100$. On the other hand, when $\omega_{pe}/\Omega_{ce}$ is reduced to 3 [panel (C)], one begins to see that the diffusion along $V_\perp$ space is restricted. The trend is further enhanced when $\omega_{pe}/\Omega_{ce}$ is reduced even further to 1. In Panel (D), the electron beam VDF shows that the diffusion along $V_\perp$ is markedly limited when compared with the previous cases. The reason for such a difference may be understood from the velocity diffusion coefficient for magnetized case. When one includes the $B$ field effects, then the velocity diffusion coefficient includes the factor
\begin{equation}
\sum_{n=-\infty}^\infty J_n^2\left(\frac{k_\perp V_\perp}{\Omega_{ce}}\right)
\delta(\omega-k_\parallel V_\parallel-n\Omega_{ce}).
\label{bessel}\end{equation}
For sufficiently strong magnetic field, the diffusion along $V_\perp$ space will be severely limited by the Bessel function factor, since $J_n^2(k_\perp V_\perp/\Omega_{ce})$ decreases in magnitude as $V_\perp$ increases. This may explain the limited diffusion characteristics along $V_\perp$ in a qualitative sense, but in order to fully characterize the velocity space diffusion behavior, one must actually solve the diffusion equation, which is beyond the scope of the present paper.

\begin{figure}
\centering
\includegraphics[width=0.75\textwidth]{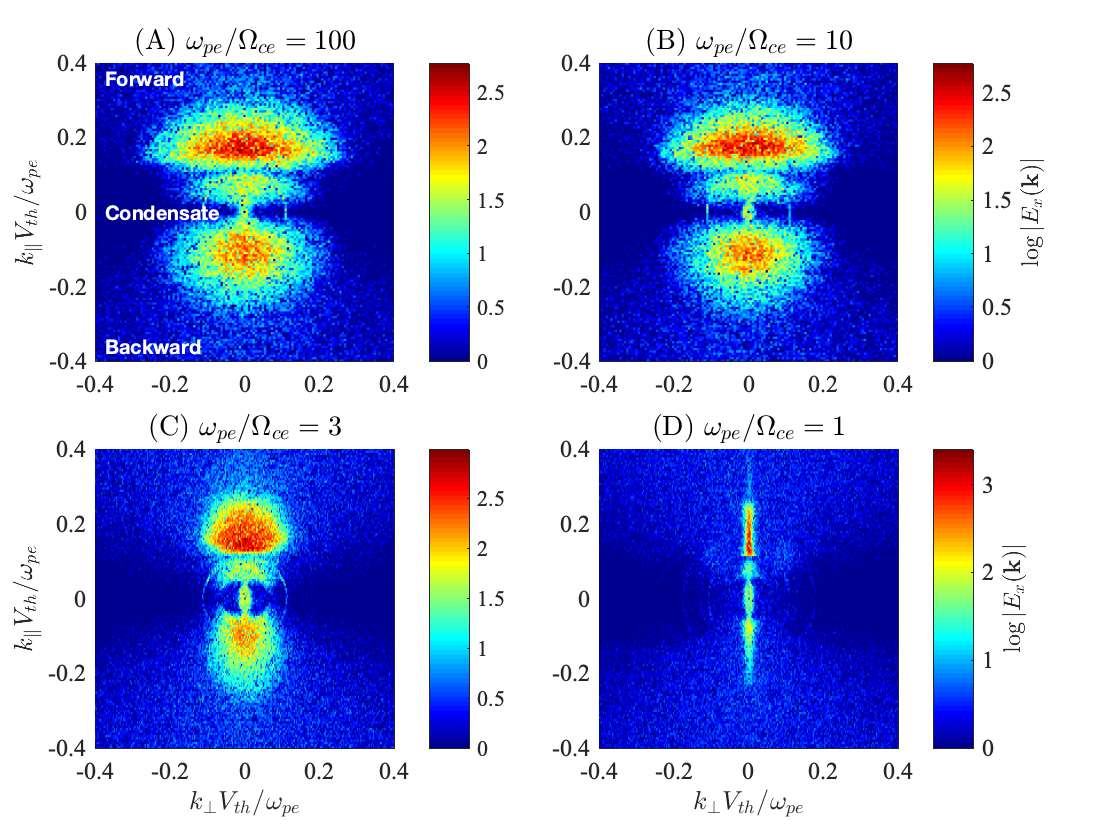}
\caption{Longitudinal mode spectra, $|E_x({\bf k})|$, at $\omega_{pe}t=2000$ for (A) $\omega_{pe}/\Omega_{ce}=100$; (b) $\omega_{pe}/\Omega_{ce}=10$; (C) $\omega_{pe}/\Omega_{ce}=3$; and (D) $\omega_{pe}/\Omega_{ce}=1$.}
\label{F2}
\end{figure}

Figure \ref{F2} displays the longitudinal electric field spectrum in 2D space, $|E_x({\bf k})|$, in logarithmic colormap scale. Positive $k_\parallel$ corresponds to parallel wave vector component along the beam propagation direction (Forward propagation), while $k_\parallel<0$ corresponds to the anti-parallel direction (Backward propagation). $k_\perp$ denotes the wave number perpendicular to the beam propagation direction (and also perpendicular to the ambient magnetic field). For case (A) [$\omega_{pe}/\Omega_{ce}=100$] the spectral pattern is virtually identical to the unmagnetized case \citep{Lee19} in that, the forward-propagating (or $k_\parallel>0$) mode, which can be interpreted as quasi Langmuir waves (here, we use the term ``quasi'' since for magnetized plasmas, the high-frequency longitudinal mode is Langmuir wave-like for quasi-parallel propagation direction, but gradually turns into the upper-hybrid wave for oblique to quasi-perpendicular angles of propagation), has undergone a significant back-scattering via combined decay and induced scattering by the time the beam-plasma instability has progressed to $\omega_{pe}t=2000$. The backward-propagating waves are those corresponding to $k_\parallel<0$. Also notice the long wavelength modes broadly distributed near $k\sim0$, which is due to the Langmuir condensation effects. For case (B), which corresponds to $\omega_{pe}/\Omega_{ce}=10$, the spectrum is qualitatively similar to panel (A). Recall that the electron VDFs for (A) and (B) are also quite similar to the unmagnetized case. Thus, the situation with the spectra is consistent with that of the electrons.

For case (C), for which $\omega_{pe}/\Omega_{ce}=3$, the longitudinal electric field spectrum appears to be shrunk along $k_\perp$ direction. Again, this can be understood in a qualitative sense. As with the particle diffusion coefficient, the linear growth rate of the longitudinal mode also has the same Bessel function factor \eqref{bessel}. As such, for sufficiently high $k_\perp$, the factor $J_n^2(k_\perp V_\perp/\Omega_{ce})$ decreases in magnitude, thus preventing the wave growth at high $k_\perp$ regime. Such a feature is also reflected in the backscattered ($k_\parallel<0$) mode as well as the condensate ($k\sim0$) mode. In spite of such a qualitative explanation, the actual reconstruction of the wave spectrum requires the numerical solution of underlying equations of weak turbulence theory in magnetized plasmas, which is not available yet. For case (D) with $\omega_{pe}/\Omega_{ce}=1$, the wave dynamics in both linear and nonlinear stages appear almost one dimensional. For strongly magnetized plasmas the electron gyro radius around the ambient magnetic field can be quite small while the field-aligned motion is much less restricted. As a result, the wave growth, saturation, and nonlinear mode coupling all appear to be confined to quasi 1D space. Again, to fully characterize this one must actually solve the fundamental equations of magnetized plasma weak turbulence theory.

\begin{figure}
\centering
\includegraphics[width=0.75\textwidth]{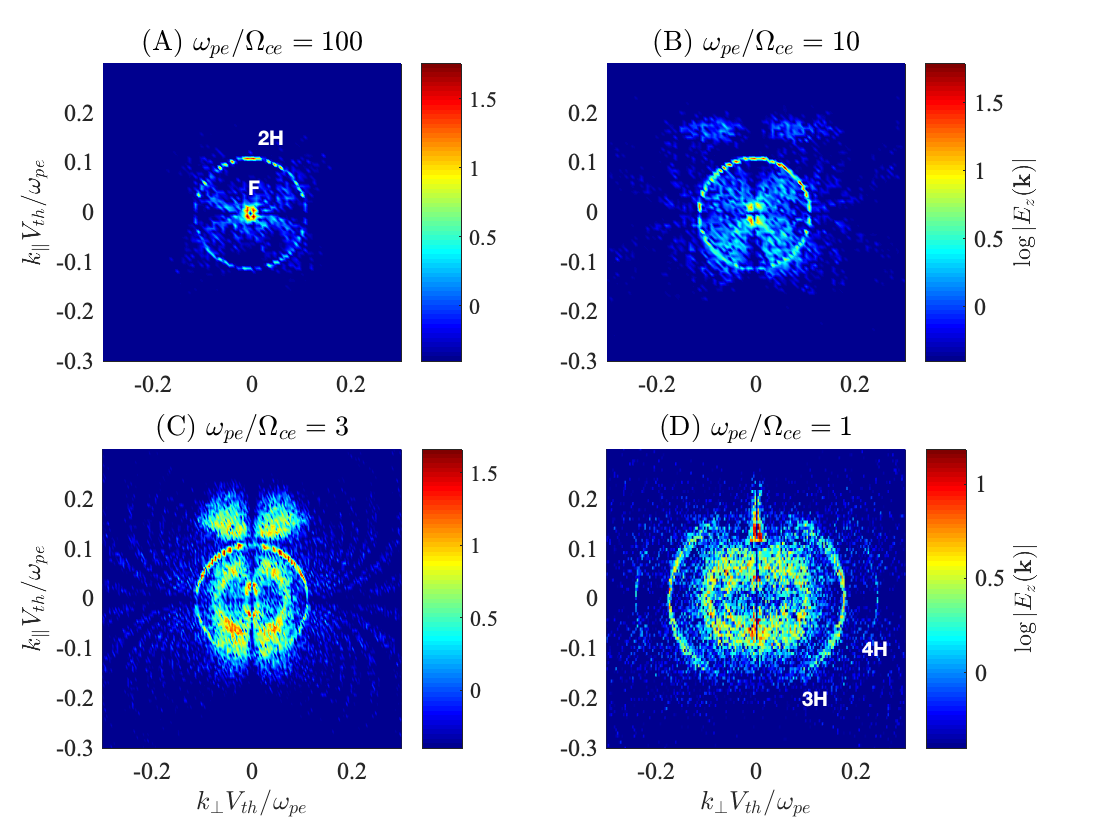}
\caption{Transverse mode spectra, $|E_z({\bf k})|$, at $\omega_{pe}t=2000$ for (A) $\omega_{pe}/\Omega_{ce}=100$; (b) $\omega_{pe}/\Omega_{ce}=10$; (C) $\omega_{pe}/\Omega_{ce}=3$; and (D) $\omega_{pe}/\Omega_{ce}=1$.}
\label{F3}
\end{figure}

Figure \ref{F3} plots the wave spectra for the same four cases except that the electric field spectra correspond to the transverse polarization, $|E_z({\bf k})|$. In the simulation it is not so easy to separate the radiation modes from transverse-polarized modes associated with non-escaping plasma normal modes. In the cold magnetized plasma the extraordinary ($X$) and ordinary ($O$) modes are transverse radiations that can escape to free space in an inhomogeneous medium. On the other hand, the slow-extraordinary, or the $Z$ mode, and the whistler, or $W$, mode are normal modes trapped within the plasma that cannot escape to free space. The transverse wave spectra shown in Figure \ref{F3} are projections of all these modes, regardless of the characteristic wave frequency associated with each mode, onto 2D wave number space. As such, the interpretation of the spectra requires some further analyses, which we will get to shortly. For the moment, paying attention to panel (A), the most prominent feature are the concentric circles (the radius of the central circle is so small that the inner circle appears as a dot). These correspond to the fundamental and harmonic plasma emissions, with the inner circle being the fundamental emission with $\omega\sim\omega_{pe}$, while the outer circle depicts the second harmonic emission with $\omega\sim2\omega_{pe}$. Such a concentric ring-like spectral feature is highly reminiscent of the unmagnetized case \citep{Lee19}. Note that the circular plasma emission pattern is consistently present in all four cases, and in fact, the intensity becomes slightly higher as $\omega_{pe}/\Omega_{ce}$ decreases in magnitude. For case (D), in particular, the third harmonic ($\omega\sim3\omega_{pe}$) emission is also present, and even what appears to be a faint fourth harmonic. Moreover, the ring associated with the second-harmonic emission has increased in width. This has to do with the modification of the dispersion relation in the presence of a strong magnetic field, which we will come back to later. Returning to case (A), one may notice that a faint quadrupole pattern of enhanced wave intensity is present, which seems to connect the two plasma emission inner and outer rings. This superposed pattern becomes more evident in panel (B), as $\omega_{pe}/\Omega_{ce}$ is reduced to 10. For $\omega_{pe}/\Omega_{ce}=3$ [panel (C)], the quadrupole pattern gets not only distorted but also, a new enhanced wave intensity appears outside the outer ring of second-harmonic emission (the rabbit ear-like protrusion). This was also present, albeit very weakly, in panel (B), but now it becomes quite evident. We will investigate this in more detail later. For the fourth case of strongly magnetized plasma [case (D) with $\omega_{pe}/\Omega_{ce}=1$], the quadrupole pattern between the fundamental emission (inner dot) and second-harmonic emission (outer ring) is now more or less merged with the second-harmonic outer ring to the extent that it is difficult to separate the two. Notice also a narrow jet-like feature, which is present along positive $k_\parallel$ axis. This is probably the rabbit ear-like protrusion that is apparent in case (C), which has now collapsed into a narrow quasi 1D structure. A salient feature associated with the transverse spectrum is that regardless of the strength of the background magnetic field, the effects of Bessel function factor are not apparent. That is, for the velocity distribution and the longitudinal mode, the increasing value of $\omega_{pe}/\Omega_{ce}$ has the effect of restricting the perpendicular dynamics so that both the velocity distribution and the longitudinal mode spectrum had tendencies to become quasi one-dimensional. In contrast, the perpendicular mode structure associated with transverse mode spectrum remains unchanged. This seems to imply that the transverse mode generation is unaffected by the Bessel function factor $J_n^2(k_\perp V_\perp/\Omega_{ce})$, which in turn, could mean that the transverse mode generation processes are inherently nonlinear.

\begin{figure}
\centering
\includegraphics[width=0.75\textwidth]{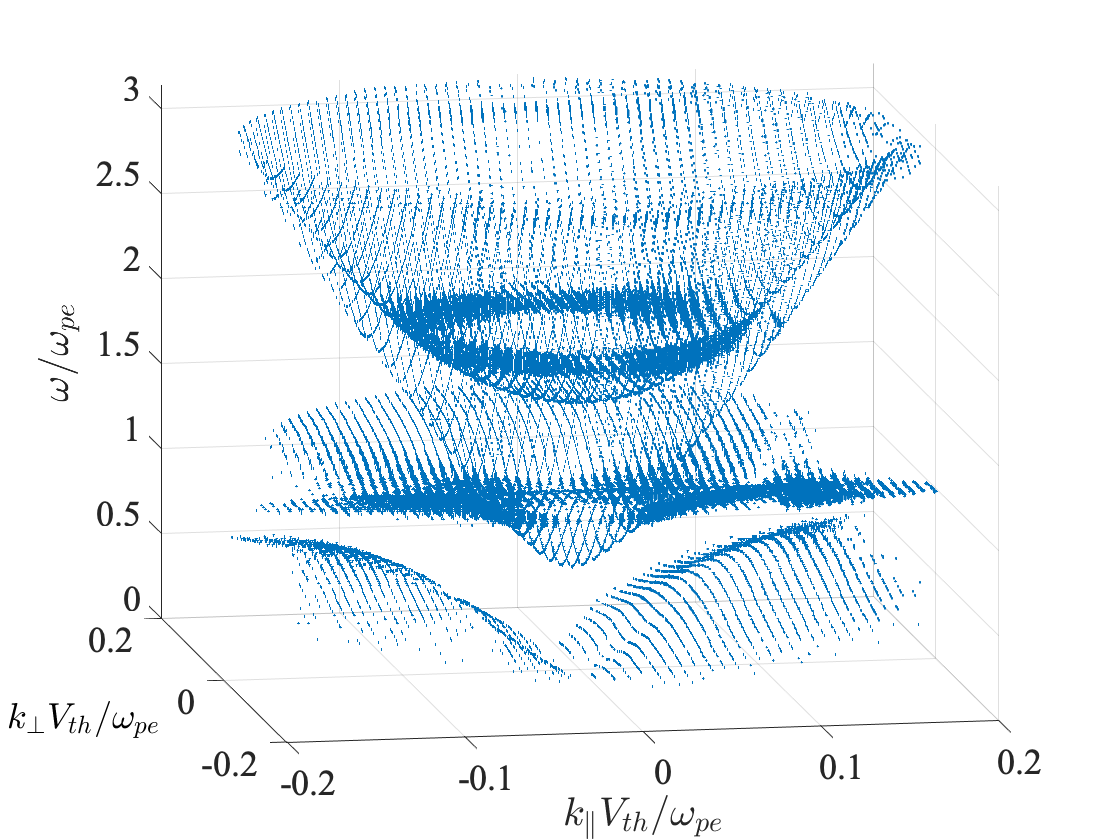}
\caption{Scatter plot of the wave spectrum for strongly magnetized case of $\omega_{pe}/\Omega_{ce}=1$.}
\label{F4}
\end{figure}

The complex behavior and the morphology associated with the transverse wave spectrum as shown in terms of the projections onto 2D ${\bf k}$ space can only provide a limited understanding. In order to further analyze the situation we consider the full three-dimensional structure associated with the transverse mode spectrum. Thus, in the remainder of the present paper, we revisit certain aspects associated with the wave spectra that the simple projection onto 2D wave number space could not elucidate. Let us first pay attention to panel (D) of Figure \ref{F3}, where we have noted that the second-harmonic ring-like spectrum has an increased width and that the intermediate intensity that connects the inner ring and outer ring appeared to merge with the outer ring. These features are the result of 2D projection. In Figure \ref{F4}, we plot the 3D spectrum including the characteristic frequency associated with each point in 2D ${\bf k}$ spectrum. We accomplish this by plotting the wave spectral intensity as a three-dimensional scatter plot in $(k_\parallel,k_\perp,\omega)$ space. As Figure \ref{F4} makes it evident, the second-harmonic ring is associated with the $X$ mode. For $\omega_{pe}/\Omega_{ce}=1$, the $X$ and $O$ mode are clearly separated, $X$ mode dispersion surface being situated above the $O$ mode surface. The $O$ and $Z$ mode surfaces are intertwined near the cutoff frequencies ($k\sim0$) so that the two modes undergo switchovers. The low-frequency $W$ mode surface lies underneath the other three dispersion surfaces. The modification and broadening of $X$ mode dispersion surface near the cutoff leads to the increase of width associated with the harmonic emission spectral ring.

It is quite clear from Figure \ref{F4} that the $Z$ mode intensity is rather high, which obscures the second-harmonic ring from the $Z$ mode wave intensity when projected onto 2D ${\bf k}$ space. However, when the spectrum is separated in frequency one may glean that the second-harmonic ring and the broad $Z$ mode spectrum belong to separate branches. Note that the jet-like enhanced wave intensity along positive $k_\parallel$ axis (the jet-like protrusion) is associated with the $Z$ mode, which the present viewing angle makes it difficult to visualize, but when plotted with different viewing angle, such a feature becomes more evident (not shown).

For the present case of $\omega_{pe}/\Omega_{ce}=1$, the radiation emission at the fundamental plasma frequency is in the sense of $O$ mode, but since the $O$ mode and $Z$ mode exist in close vicinity of each other in both frequency and wave number, it is difficult to determine which mode is which. Judging from the enhanced $W$ mode intensity, the radiation emission at the fundamental plasma frequency might be mediated by decay processes involving $W$ modes, as discussed by \citet{Chiu, Chin, Ni20, Ni21} in spite of the negative assessment by \citet{Melrose75}. Note that we have imposed the upper limit on the frequency at $3\omega_{pe}$ so that fourth harmonic is not shown, but one can see, albeit with some difficulties, that the transverse mode intensity is somewhat enhanced in the vicinity of $3\omega_{pe}$. However, it is difficult to visually discern whether the third-harmonic emission is in the sense of $X$ or $O$ mode polarization due to the fact that at such a frequency the $X$ and $O$ mode dispersion surfaces are in close proximity, and also because of the fact that the third-harmonic emission is quite weak in intensity.

The complex features associated with the various modes necessitate further theoretical analysis, which is the subject of future research. Even with the 3D spectral scatter plot, the best we can do is to provide verbal descriptions based on visual examination. However, with a full development of weak turbulence theory in magnetized plasmas, the future goal is to replicate these various features by theoretical means. Only then could one truly understand the effects of ambient magnetic field on plasma emission process.

\begin{figure}
\centering
\includegraphics[width=0.75\textwidth]{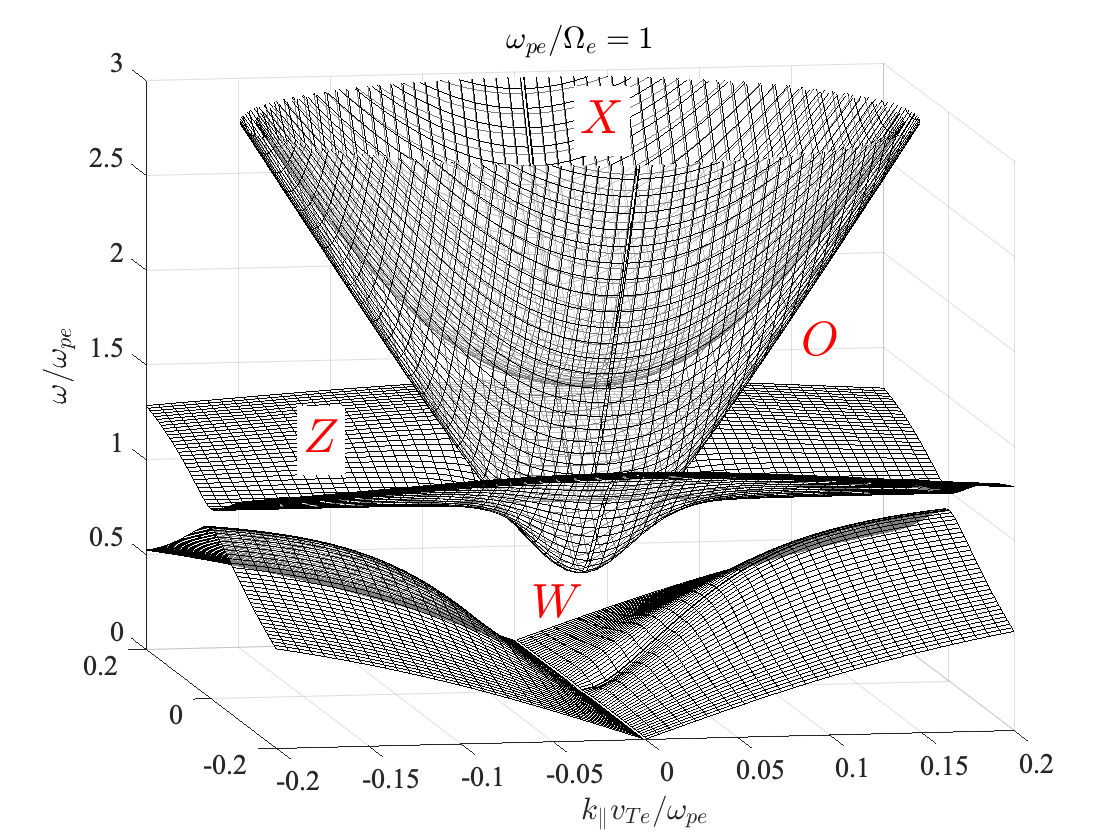}
\caption{Magneto-ionic dispersion surfaces for $\omega_{pe}/\Omega_{ce}=1$.}
\label{F5}
\end{figure}

It is interesting to note that the simulated wave spectrum, when plotted as a three-dimensional scatter plot, traces the magneto-ionic dispersion surfaces rather accurately. To see this, we have plotted in Figure \ref{F5}, the cold-plasma mode dispersion relations as four separate manifolds (or surfaces) in $\omega$ versus $(k_\perp,k_\parallel)$ space. The nested surfaces are, from top, the fast extraordinary (or $X$) mode, the ordinary (or $O$) mode, which is coupled with the slow extraordinary (or $Z$) mode near the plasma frequency $\omega_{pe}$ or upper-hybrid frequency $\omega_{uh}=(\omega_{pe}^2+\Omega_{ce}^2)^{1/2}$ regime, and underneath these is the whistler (or $W$) mode dispersion surface. Upon comparison with the 3D scatter plot of simulated wave intensity, it becomes quite evident that the magneto-ionic theory of waves is quite valid as far as the mode configuration in $\omega$ versus ${\bf k}$ space is concerned. That is, the weakly turbulent processes taking place in strongly magnetized plasma ($\omega_{pe}/\Omega_{ce}=1$) seem to involve linear and nonlinear wave-particle and wave-wave interactions among particles and linear eigen-modes of the magnetized plasma, but the eigen-modes themselves are well described by the linear wave theory. Note that for warm plasmas, besides the cold magneto-ionic modes, thermal acoustic (or sonic) modes are also excited, but as such modes are low frequency modes in the vicinity of ion plasma or ion cyclotron frequency, it is difficult to set the genuine thermal sonic modes apart from numerical noise-like fluctuations according to the present PIC simulation. Recall that the traditional approach to interpreting plasma emission in magnetized plasmas had been to invoke these thermal sonic modes as being important rather than $W$ mode \citep{Trakhtengerts, MelroseSy, Melrose78, Melrose80, Zlotnik, Willes}.

\begin{figure}
\centering
\includegraphics[width=0.75\textwidth]{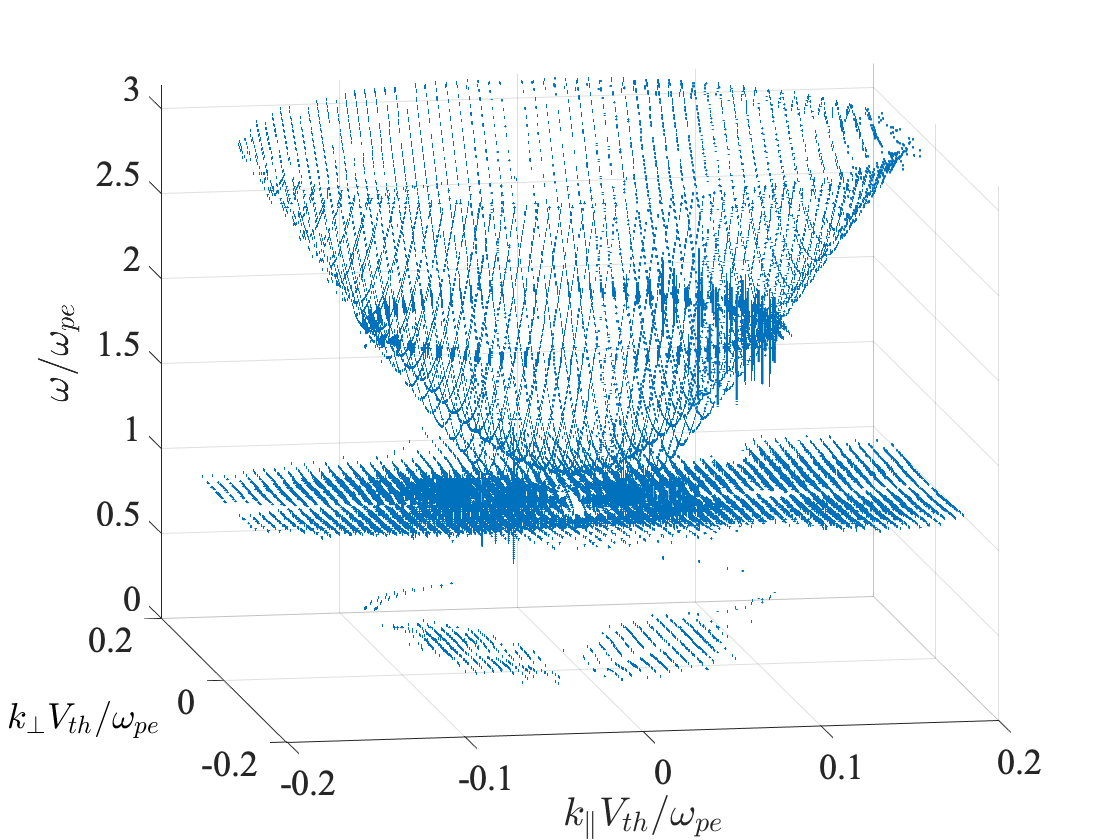}
\caption{Scatter plot of the wave spectrum for moderately magnetized case of $\omega_{pe}/\Omega_{ce}=3$.}
\label{F6}
\end{figure}

We repeat the 3D scatter plot of the wave spectrum for case (C) [$\omega_{pe}/\Omega_{ce}=3$], which is plotted in Figure \ref{F6}. In this case, the $X$ and $O$ mode dispersion surfaces are very close to each other at $\omega\sim2\omega_{pe}$. As such, the harmonic emission, which is seen as the ring-like enhancement, might have a mixed $X$/$O$ mode polarization, but the visual inspection is insufficient to ascertain this. It is interesting to note that the fundamental emission is buried within a broad swath of enhanced $Z$ mode spectrum, which makes it difficult to distinguish one from the other. The peculiar feature in Figure \ref{F3}D, namely, the quadrupole pattern of enhanced intensity in between the inner and outer ring as well as the pair of rabbit ear-like protrusion outside the outer ring, can now reasonably be identified with the $Z$ mode wave intensity. When the spectrum is projected onto 2D space, one cannot set part all these features from each other, but separating the spectrum in $\omega$ space facilitates the delineation of different modes. Note that the low-frequency $W$ mode is associated with a weak but visible wave intensity, which again points to the cold-plasma $W$ mode participating in the plasma emission \citep{Chiu, Chin, Ni20, Ni21}. Again, even lower-frequency thermal sonic modes could not be distinguished from numerical noise in a meaningful way in the present simulation.

\begin{figure}
\centering
\includegraphics[width=0.75\textwidth]{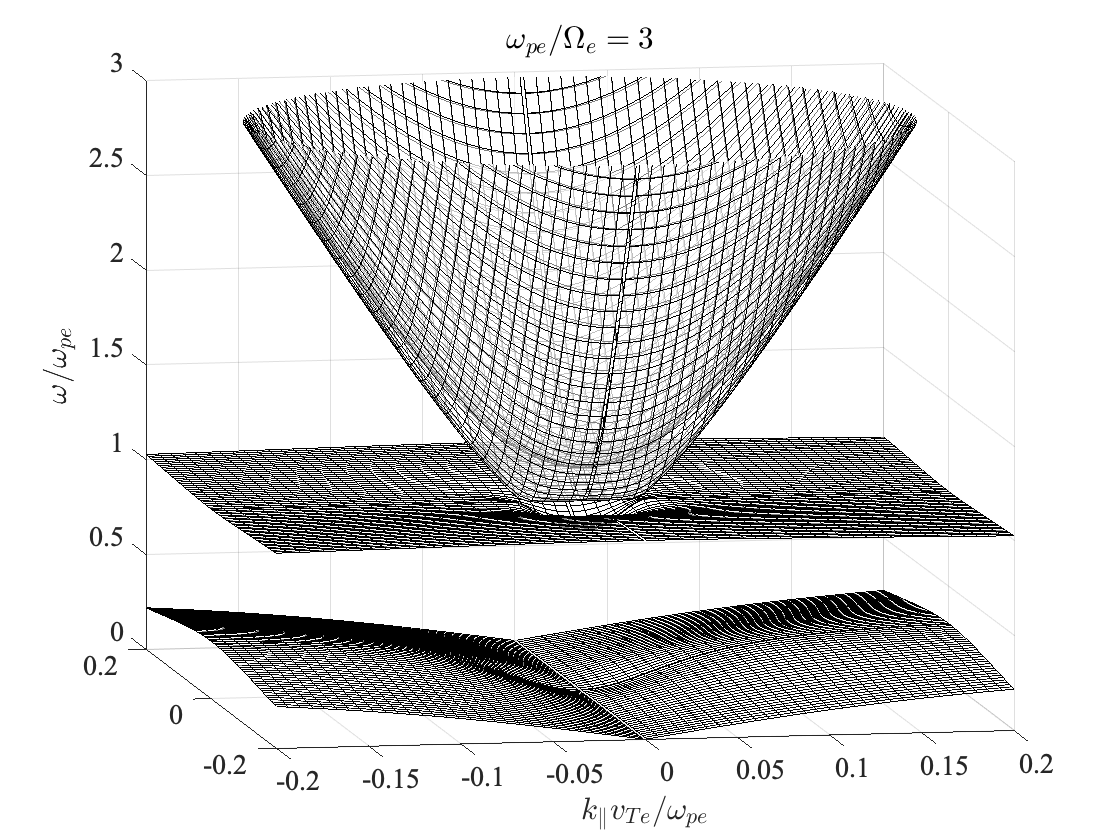}
\caption{Magneto-ionic dispersion surfaces for $\omega_{pe}/\Omega_{ce}=3$.}
\label{F7}
\end{figure}

We have not repeated the same scatter plots for cases (A) and (B), but shown in Figure \ref{F7} is the analogue of Figure \ref{F5} except that the magneto-ionic dispersion surfaces correspond to $\omega_{pe}/\Omega_{ce}=3$ are plotted. The comparison between the scatter plot of wave intensity in Figure \ref{F6} and the magneto-ionic dispersion relation again confirms that the basic eigen-modes excited in the simulation are those that can be described by the linear theory of plasma waves. However, the linear plasma modes undergo linear and nonlinear interactions among themselves as well as with the particles. This is the basic picture implicit in the weak turbulence theory, which shows that the plasma emission problem is not only relevant for solar radio bursts phenomena but is also an ideal problem to test and validate the weak turbulence theory.

\begin{figure}
\centering
\includegraphics[width=0.75\textwidth]{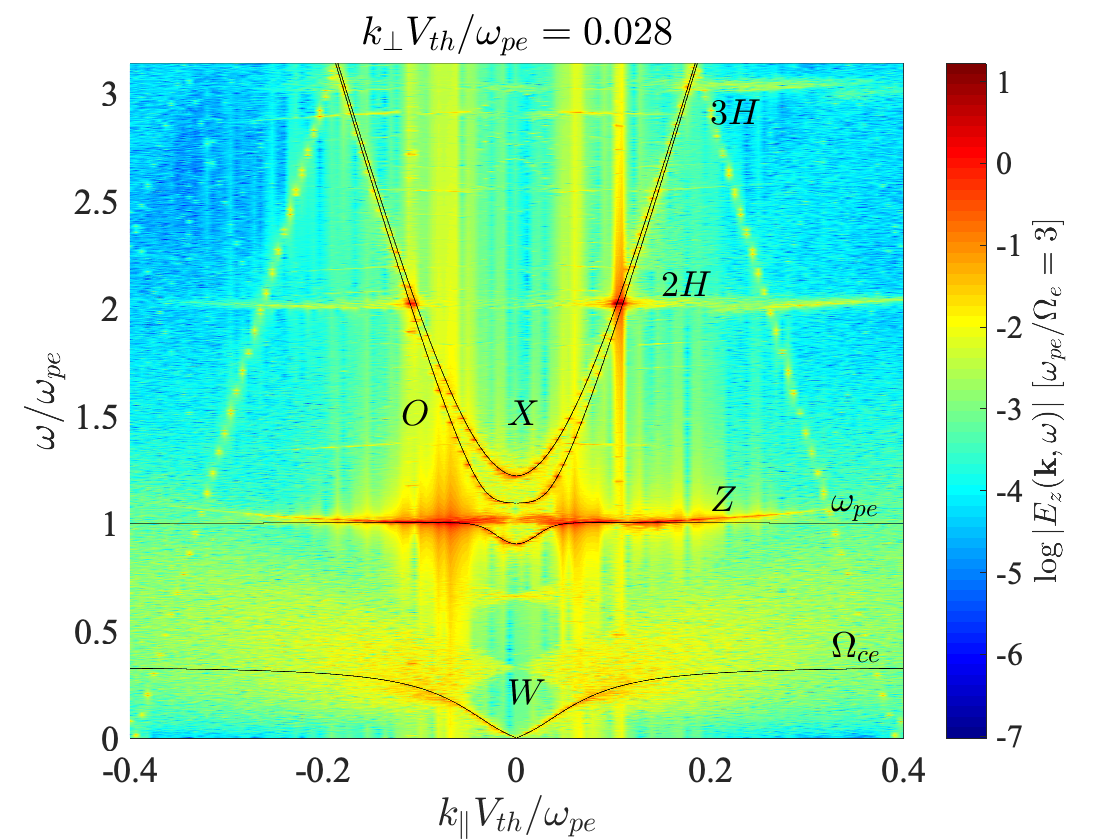}
\caption{The two-dimensional section of the simulated transverse wave spectrum for $\omega_{pe}/\Omega_{ce}=3$. Think black curves depict the magneto-ionic dispersion relation.}
\label{F8}
\end{figure}

Despite the good agreement between Figures \ref{F6} and \ref{F7}, we note that both the scatter plot and the dispersion surface plot obscure the complicated mode structure near the $X$ and $O$ mode cutoff regions (near $k\sim0$). In order to furnish additional information on the mode structure near $k\sim0$ region, we plot in Figure \ref{F8}, the two-dimensional cut associated with the wave spectrum in $\omega$ versus $k_\parallel$ space, for the case of $\omega_{pe}/\Omega_{ce}=3$ [case (C)], which is generated for fixed $k_\perp V_{th}/\omega_{pe}=0.028$. The simulated transverse electric field energy spectrum is shown in colormap. Superposed with thin curves are the theoretical dispersion relations. As one may appreciate, the theoretical magneto-ionic dispersion curves fit almost perfectly with the simulated spectrum. Note the well-defined second-harmonic emission near $2\omega_{pe}$ and a faint but identifiable third-harmonic emission at $3\omega_{pe}$. Near the fundamental plasma frequency, however, the enhanced wave spectrum has a complex frequency structure. Both $X$ and $O$ modes have enhanced intensity, which means the fundamental emission for type III radio sources characterized by $\omega_{pe}/\Omega_{ce}=3$ could have a mixed polarization. The second-harmonic emission could also have a mixed polarization, but as noted before, because of the close proximity of the two dispersion curves, it is difficult to distinguish the two modes. Near $\omega\sim\omega_{pe}$, the $Z$ mode has the high intensity, and the enhanced $Z$ mode spectrum extends beyond the infinitely narrow normal mode dispersion curve. As noted above, the $W$ mode shows a slight enhanced intensity, which could be the result of decay instability involving the beam-generated $Z$ mode, or the electron beam directly exciting the $W$ mode by linear instability mechanism. Again, all of these multi-step processes must be investigated by a first-principle theory in the future.

\section{Summary and Conclusions}\label{sec4}

To summarize and conclude the present paper, we have carried out a series of particle-in-cell (PIC) simulations of the plasma emission process in magnetized plasma. The plasma emission, or the emission of radiation at the fundamental plasma frequency and/or its harmonics, is widely believed to be the essential radiation emission mechanism responsible for the solar coronal and interplanetary type II and type III radio bursts. The plasma emission problem is also an ideal platform to test and validate the plasma weak turbulence from the perspective of fundamental plasma physics. In the literature the plasma emission problem is studied by means of theory and simulations, but almost invariably, the effects of ambient magnetic field are ignored. In contrast, in the present paper, we have considered the effects of background magnetic field by systematically varying the frequency ratio, $\omega_{pe}/\Omega_{ce}$. When $\omega_{pe}^2/\Omega_{ce}^2=4\pi n_0m_ec^2/B_0^2$ is high, then the plasma can be considered as weakly magnetized. If this ratio is low, on the other hand, then the plasma is strongly magnetized. We found that $\omega_{pe}/\Omega_{ce}$ higher than 100 (or even as low as 10) -- for which $\omega_{pe}^2/\Omega_{ce}^2$ is $10^4$ and $100$, respectively, can be considered as virtually unmagnetized. Thus, the standard weak turbulence theory developed under the unmagnetized plasma approximation may be applicable. However, when we lower the ratio to $\omega_{pe}/\Omega_{ce}=3$ (or $\omega_{pe}^2/\Omega_{ce}^2=9$), we begin to see that the underlying dynamical processes markedly deviate from those of unmagnetized plasma. We employed a number of diagnostic tools in order to delineate and comprehend various features associated with the wave spectrum for $\omega_{pe}/\Omega_{ce}=3$ as well as for even lower value of $\omega_{pe}/\Omega_{ce}=1$, and found that the presence of strong ambient magnetic field restricts the particle diffusion along perpendicular velocity and the generation of longitudinal mode along perpendicular wave number. On the other hand, the transverse modes are unaffected by the presence of background magnetic field as far as the range of perpendicular wave number is concerned, but the transverse spectrum generated in a magnetized plasma exhibits rather complex pattern. In the present paper we have provided verbal descriptions of the simulation results with the aid of sophisticated diagnostic techniques, but in order to fully comprehend the underlying physical processes, one must formulate and solve the basic equations of weak turbulence theory in magnetized plasmas, which must be done in the future. The present simulation results, however, will be valuable in that our work will provide a constraint against which, such future theories could be tested and validated.

\acknowledgments

S.L. acknowledges ...
...
P.H.Y. acknowledges NASA Grant 80NSSC19K0827 and
NSF Grant 1842643 to the University of Maryland.

\bibliography{Lee+21_ApJ}

\end{document}